\documentclass[aps,prc,twocolumn,showpacs,amsmath,amssymb]{revtex4}
\usepackage{ulem}
\usepackage{graphicx}
\usepackage{epstopdf}
\usepackage{dcolumn}
\usepackage{bm}
\usepackage{hyperref}
\usepackage{multirow}
\usepackage{rotating}
\usepackage{url}
\usepackage{ifthen}\newboolean{2column}\setboolean{2column}{true}

\begin{document}


\title{Nuclear fourth-order symmetry energy and its effects on neutron star properties in the relativistic Hartree-Fock theory}

\author{Zhi Wei Liu}
\affiliation
{School of Nuclear Science and Technology, Lanzhou University, Lanzhou 730000, China}

\author{Zhuang Qian}
\affiliation
{School of Nuclear Science and Technology, Lanzhou University, Lanzhou 730000, China}

\author{Ruo Yu Xing}
\affiliation
{School of Nuclear Science and Technology, Lanzhou University, Lanzhou 730000, China}

\author{Jia Rui Niu}
\affiliation
{School of Nuclear Science and Technology, Lanzhou University, Lanzhou 730000, China}

\author{Bao Yuan Sun \footnote{sunby@lzu.edu.cn}}
\affiliation
{School of Nuclear Science and Technology, Lanzhou University, Lanzhou 730000, China}

\date{\today}

\begin{abstract}
Adopting the density dependent relativistic mean-field (RMF) and relativistic Hartree-Fock (RHF) approaches, the properties of the nuclear fourth-order symmetry energy $S_4$ are studied within the covariant density functional (CDF) theory. It is found that the fourth-order symmetry energies are suppressed in RHF at both saturation and supranuclear densities, where the extra contribution from the Fock terms is demonstrated, specifically via the isoscalar meson-nucleon coupling channels. The reservation of $S_4$ and higher-order symmetry energies in the nuclear equation of state then affects essentially the prediction of neutron star properties, which is illustrated in the quantities such as the proton fraction, the core-crust transition density as well as the fraction of crustal moment of inertia. Since the Fock terms enhance the density dependence of the thermodynamical potential, the RHF calculations predict systematically smaller values of density, proton fraction and pressure at the core-crust transition boundary of neutron stars than density dependent RMF ones. In addition, a linear anti-correlation between the core-crust transition density $\rho_t$ and the density slope of symmetry energy $L$ is found which is then utilized to constrain the core-crust transition density as $\rho_t\thicksim[0.069, 0.098]~\rm{fm}^{-3}$ with the recent empirical information on $L$. The study clarifies the important role of the fourth-order symmetry energy in determining the properties of nuclear matter at extreme isospin or density conditions.
\end{abstract}

\maketitle

\section{Introduction}
The study on nuclear equation of state (EoS), especially its properties at extreme conditions, is not only a longstanding goal in nuclear related science but in astrophysics \cite{Lattimer2000PR333.121,Steiner2005PR411.325,Baran2005PR410.335,Li2008PR464.113}. In recent years, facilities for radioactive ion beam (RIB) are developing competitively around the terrestrial laboratories, which have made great progress in exploring the nuclear EoS at both supranuclear and subnuclear densities \cite{CSR.HIRFL, RIB.RIKEN, FAIR, FRIB, Balantekin2014MPLA29.1430010}. In particular, the isospin-asymmetric part of the EoS, namely the nuclear symmetry energy, is proved to be a crucial issue in understanding the physics of several terrestrial experiments and astrophysical observations, such as neutron skin thickness \cite{Brown2000PRL85.5296, Centelles2009PRL102.122502}, dipole excitation modes of stable or exotic nuclei \cite{Tamii2011PRL107.062502}, isospin diffusion and $\pi^+/\pi^-$ ratio in heavy-ion collisions at intermediate energies \cite{Chen2005PRL94.032701, Li2008PR464.113, Xiao2009PRL102.062502}, parity violating electron scattering \cite{Roca2011PRL106.252501}, as well as the radius, the moment of inertia, the stability of matter and cooling mechanism of neutron stars \cite{Horowitz2001PRL86.5647, Horowitz2002PRC66.055803, Fattoyev2011PRC84.064302, Worley2008AJ685.390, Kubis2007PRC76.025801, Newton2013AJL779.L4}.

Theoretically, the nuclear symmetry energy is introduced by expanding the binding energy per nucleon in a Taylor series with respect to the isospin asymmetry, and is usually approximated to its second-order term $S_2$ for convenience. Although a number of phenomenological and microscopic nuclear models have been devoted to constrain the symmetry energy around the saturation density $\rho_0$ and its density dependence \cite{Akmal1998PRC58.1804, Chen2005PRC72.064309, Dalen2005PRC72.065803, Chen2007PRC76.054316, Li2008PRC78.028801, Taranto2013PRC87.045803, Lopes2014PRC89.025805}, large uncertainties still remains \cite{Tsang2012PRC86.015803, Lattimer2013AJ771.51, Li2013PLB727.276, Lattimer2014EPJA50.40}. Recently, a data collective analysis constrained the symmetry energy $S_2(\rho_0)$ at the saturation density as $S_2(\rho_0)=31.7\pm3.2$~MeV, and its density slope parameter as $L=58.7\pm28.1$~MeV retaining a relatively large error bar \cite{Oertel2017RMP89.015007}, implying the necessity to improve the work further in both experimental and theoretical sides.

Recently, it has been indicated that the higher-order terms of symmetry energy than $S_2$ may become non-negligible and should be considered carefully under some extreme physical conditions \cite{Zhang2001CPL18.142, Steiner2006PRC74.045808, Xu2009PRC79.035802, Xu2009AJ697.1549, Cai2012PRC85.024302, Seif2014PRC89.028801, Kaiser2015PRC91.065201, Wang2017PLB773.62, Pu2017arXiv1708.02132, Gonzalez2017arXiv1706.02736}. Especially, the inclusion of the fourth-order term of nuclear symmetry energy $S_4$ beyond the parabolic approximation\cite{Bombaci1991PRC44.1892} in the EoS could exert great impact on the description of neutron star properties, such as the core-crust transition density \cite{Xu2009PRC79.035802, Xu2009AJ697.1549, Cai2012PRC85.024302, Gonzalez2017arXiv1706.02736}, the proton fraction and the critical density for the direct URCA (DUrca) process \cite{Zhang2001CPL18.142, Steiner2006PRC74.045808, Cai2012PRC85.024302, Seif2014PRC89.028801}. However, because of the absence of experimental information, there is still large uncertainty in constraining the magnitude of $S_4$, even at the saturation density $\rho_0$. It is found that the magnitude of $S_4(\rho_0)$ is generally smaller than 2 MeV within the non-relativistic \cite{Chen2009PRC80.014322, Pu2017arXiv1708.02132} and relativistic mean-field models \cite{Cai2012PRC85.024302}, as well as the chiral pion-nucleon dynamics \cite{Kaiser2015PRC91.065201, Wellenhofer2016PRC93.055802}. Alternatively, the analysis from quantum molecular dynamics model\cite{Nandi2016PRC94.025806} and the extraction from an extended nuclear mass formula \cite{Wang2017PLB773.62} predict significantly large values of $S_4(\rho_0)$, for example $20.0\pm4.6$~MeV in latter one. In addition, the kinetic part $\varepsilon_k$ of the energy density functional has been identified as a particularly good indicator of the short-range correlated (SRC) nucleon pairs \cite{Vidana2011PRC84.062801, Carbone2012EPL97.22001, Xu2012CPL29.122102, Xu2013JPCS420.012090, Hu2013PTEP2013.103D02, Zhang2014EPJA50.113, Hen2015PRC91.025803, Cai2016PRC93.014619}. Theoretically, this SRC pairs can happen due to the tensor part of the nucleon-nucleon interaction \cite{Sargsian2005PRC71.044615, Schiavilla2007PRL98.132501, Subedi2008Science320.1476, Hen2014Science346.614}, which consequently is revealed to reduce the kinetic part of $S_2(\rho_0)$ \cite{Carbone2012EPL97.22001, Xu2012CPL29.122102, Zhang2014EPJA50.113, Hen2015PRC91.025803, Zhao2015JPG42.095101, Cai2016PRC93.014619} and increase that of $S_4(\rho_0)$ \cite{Cai2015PRC92.011601} significantly.

The covariant density functional (CDF) theory, based on the meson exchange diagram of nuclear force, has achieved great successes in describing the bulk properties of nuclear matter, the ground state and excitation properties of finite nuclei \cite{Reinhard1989RPP52.439, Ring1996PPNP37.193, Meng1999PRC59.154, Zhou2003PRL91.262501, Chen2003CPL20.358, Vretenar2005PR409.101, Meng2006PPNP57.470, Meng2006PRC73.037303, Meng2015JPG42.093101, Liang2015PR570.1}. During the past decades, the CDF theory without Fock terms, namely the relativistic mean field (RMF) theory, occupied the CDF market with various versions of Lagrangian \cite{Boguta1977NPA292.413, Boguta1983PLB120.289, Typel1999NPA656.331, Long2004PRC69.034319, Nikolaus1992PRC46.1757, Zhao2010PRC82.054319}. Since the Fock terms are ignored in the RMF theory, the important degrees of freedom in the meson exchange diagrams, such as the pseudo-vector $\pi$-couplings, are dropped. Moreover, the nonlocal potential as well as the tensor part of nuclear force cannot be self-consistently taken into account. With increasing computer technology, the relativistic Hartree-Fock (RHF) theory \cite{Bouyssy1987PRC36.380, Bernardos1993PRC48.2665, Shi1995PRC52.144, Marcos2004JPG30.703}, also referred as the CDF theory with Fock terms, achieved success in terms of the density-dependent meson-nucleon coupling \cite{Long2006PLB640.150, Long2007PRC76.034314, Long2010PRC81.024308}. Significant improvements were obtained by the RHF theory in exploring nuclear structure \cite{Long2006PLB639.242, Lu2013PRC87.034311, Li2016PRC93.054312}, nuclear excitation and decay modes \cite{Liang2008PRL101.122502, Liang2009PRC79.064316} as well as nuclear matter and neutron star porperties \cite{Long2006PLB639.242, Long2012PRC85.025806, Sun2008PRC78.065805, Jiang2015PRC91.025802, Zhao2015JPG42.095101}.

In fact, with the inclusion of the Fock terms in the CDF theory, it is realized that isoscalar meson-nucleon coupling except for the isovector one also plays a vital role in studying the isospin properties of nuclear matter, such as the nuclear symmetry energy and the neutron-proton effective mass splitting \cite{Long2006PLB640.150, Sun2008PRC78.065805, Long2012PRC85.025806, Jiang2015PRC91.025802, Zhao2015JPG42.095101, Li2016PRC93.015803, Sun2016EPJWC117.07011}. After including the $\Lambda$ hyperons into the $\beta$-equilibrium nuclear matter, the symmetry energy at high densities is suppressed enormously due to the extra suppression effect originating from the Fock channel, leading to a relatively small predicted value of the neutron-star radius \cite{Long2012PRC85.025806}. Additionally, it was recognized that the Fock diagrams of the meson-nucleon couplings can take the important ingredient of nuclear force --- the tensor force into account naturally \cite{Jiang2015PRC91.034326}, which softens the density-dependent behavior of the symmetry energy and consequently raises the threshold density for the direct URCA process that cools the neutron star rapidly \cite{Jiang2015PRC91.025802}. Furthermore, a sizable reduction of the kinetic part of $S_2$ at the supranuclear density region is found in the RHF calculations compared to the RMF ones, regarded partly as the effect of the nuclear tensor force \cite{Zhao2015JPG42.095101}.

The studies demonstrated the Fock terms are of great importance when talking about isospin related physics in the CDF theory, and it is interesting to investigate its effects further on higher-order symmetry energy rather than $S_2$ and the corresponding influence on neutron star properties, which founds the motivation of this work. In the following, we will briefly introduce the theoretical framework of the RHF theory for nuclear matter in Section II. Then in Section III we discuss in detail the density-dependent behavior of nuclear fourth-order symmetry energy $S_4$ and its properties at saturation density within the RHF theory and the effects of $S_4$ on the neutron star properties, including the proton fraction, the core-crust transition as well as the moment of inertia utilized to describe the glitch phenomenon, are given later. Finally, a summary is given in Section IV.

\section{Theoretical Framework}
In this section, the main formalism of the CDF theory with the inclusion of the Fock terms will be briefly recalled for nuclear matter, which is then utilized to extract various order of the symmetry energy. For more details of the CDF theory especially the RHF theory for nuclear matter, we refer the reader to Refs. \cite{Bouyssy1987PRC36.380, Sun2008PRC78.065805}.

Via the meson exchange diagrams of nuclear force, the CDF theory is started from an effective Lagrangian density which can be deduced from the collaboration with the degrees of freedom of nucleon and mesons, while the photon field is ignored for uniform nuclear matter systems. Following the standard procedure \cite{Bouyssy1987PRC36.380}, the energy density functional (EDF) is then obtained by taking the expectation value of the Hamiltonian with respect to the Hartree-Fock ground state, which consists of three parts,
\begin{subequations}\label{eq:EDF}
  \begin{align}
    \varepsilon_{kin}
    &=\sum_{ps\tau}~\bar{u}(p,s,\tau)~(\boldsymbol{\gamma}\cdot\boldsymbol{p}+M)~u(p,s,\tau),\label{eq:kinetic EDF}\\
    \varepsilon_\phi^D
    &=\frac{1}{2}\sum_{p_1s_1\tau_1}\sum_{p_2s_2\tau_2}~\bar{u}(p_1,s_1,\tau_1)\bar{u}(p_2,s_2,\tau_2)~\Gamma_\phi(1,2)\nonumber\\
    &~~~~\times\frac{1}{m_\phi^2}~u(p_2,s_2,\tau_2)u(p_1,s_1,\tau_1),\label{eq:Hartree term of the potential EDF}\\
    \varepsilon_\phi^E
    &=-\frac{1}{2}\sum_{p_1s_1\tau_1}\sum_{p_2s_2\tau_2}~\bar{u}(p_1,s_1,\tau_1)\bar{u}(p_2,s_2,\tau_2)~\Gamma_\phi(1,2)\nonumber\\
    &~~~~\times\frac{1}{m_\phi^2+\boldsymbol{q}^2}~u(p_1,s_1,\tau_1)u(p_2,s_2,\tau_2),\label{eq:Fock term of the potential EDF}
  \end{align}
\end{subequations}
where $\varepsilon_{kin}$ denotes the kinetic EDF, and $\varepsilon_\phi^D$ and $\varepsilon_\phi^E$ correspond to the Hartree (direct) and Fock (exchange) terms of the potential EDF, where $\phi~=~\sigma,~\omega,~\rho,~\pi$ represents various meson-nucleon coupling and $\Gamma_\phi(1,2)$ are corresponding interaction vertices. The Dirac spinors $u(p,s,\tau)$ depend on the momentum $p$, spin $s$, and isospin $\tau$ of nucleon,
\begin{align}\label{eq:Dirac spinor}
  u(p,s,\tau)=\left(\frac{E^*+M^*}{2E^*}\right)^{1/2}\binom{1}{\frac{\boldsymbol{\sigma}\cdot\boldsymbol{p^*}}{E^*+M^*}}~\chi_s\chi_\tau.
\end{align}
Here $\chi_s$ and $\chi_\tau$ stand for the spin and isospin wave functions, respectively. The starred quantities, which obey the effective relativistic mass-energy relation $E^{*2}=M^{*2}+\boldsymbol{p}^{*2}$, are defined as:
\begin{subequations}\label{eq:starred quantities}
  \begin{align}
    M^*
    &~=~M~+~\Sigma_S(p),\label{eq:M star}\\
    \boldsymbol{p}^*
    &~=~\boldsymbol{p}~~~+~\hat{\boldsymbol{p}}\Sigma_V(p),\label{eq:p star}\\
    E^*
    &~=~E~~-~\Sigma_0(p),\label{eq:E star}
  \end{align}
\end{subequations}
where $\Sigma_S$ is the scalar self-energy, $\Sigma_0$ and $\Sigma_V$ are the time and space components of the vector self-energy, respectively, and $\hat{\boldsymbol{p}}$ is the unit vector along $\boldsymbol{p}$.

Substituting Eq. \eqref{eq:Dirac spinor} into Eq. \eqref{eq:kinetic EDF}, the kinetic EDF is then expressed as
\begin{align}\label{eq:kinetic energy}
  \varepsilon_{kin}
  =\frac{1}{\pi^2}\sum_{i=n,p}\int^{k_{F,i}}_0p^2dp\left[M\hat{M}+p\hat{P}\right],
\end{align}
where the hatted quantities are introduced by
\begin{align}\label{eq:hatted quantities}
  \hat{M}&=\frac{M^*}{E^*}, &\hat{P}&=\frac{p^*}{E^*}.
\end{align}
$\varepsilon_{kin}$ can be divided again according to the self-energy so as to study the influence of various meson-nucleon coupling channels quantitatively \cite{Zhao2015JPG42.095101},
\begin{align}\label{eq:decomposed kinetic energy}
  \varepsilon_{kin}
  &=\frac{1}{\pi^2}\sum_{i=n,p}\int^{k_{F,i}}_0p^2dp\frac{1}{E^*}\nonumber\\
  &\times\left[M^2+p^2
  +\sum_\phi M\Sigma_S^{D,\phi}+\sum_\phi\left(M\Sigma_S^{E,\phi}+p\Sigma_V^{E,\phi}\right)\right]\nonumber\\
  &\equiv~\varepsilon_{kin}^M~+~\varepsilon_{kin}^p~+~\varepsilon_{kin}^D~+~\varepsilon_{kin}^E,
\end{align}
where $\varepsilon_{kin}^M$ and $\varepsilon_{kin}^p$ correspond to the contributions from the rest mass and the momentum, respectively, $\varepsilon_{kin}^D$ denotes the contribution from the Hartree (direct) terms of the scalar self-energy $\Sigma_S^{D,\phi}$, and $\varepsilon_{kin}^E$ represents the total contribution from the Fock (exchange) terms of the scalar self-energy  $\Sigma_S^{E,\phi}$ and the space component of the vector self-energy $\Sigma_V^{E,\phi}$.

The EoS of asymmetric nuclear matter at zero temperature is defined by its binding energy per nucleon $E(\rho,\delta)$, where $\rho=\rho_n+\rho_p$ denotes the baryon density, and $\delta\equiv(\rho_n-\rho_p)/(\rho_n+\rho_p)$ is the isospin asymmetry with $\rho_{n/p}$ the neutron/proton density. Conventionally, due to the difficulty of analytical extraction, the various order of nuclear symmetry energies can be approximately extracted by expanding the zero-temperature EoS in a Taylor series with respect to the $\delta$. The convergence of such an isospin-asymmetry expansion has been acknowledged in the self-consistent mean-field calculations, such as the CDF theory (at the first Hartree-Fock level) adopted in this work, but broken overall when second-order perturbative contributions are involved in many-body theory \cite{Wellenhofer2016PRC93.055802}. Within this approximation, the EoS is then expressed as
\begin{subequations}\label{eq:EoS}
  \begin{align}
    E(\rho,\delta)
    &~=~E_0(\rho)~+~\sum_{n=1}^\infty S_{2n}(\rho)\delta^{2n},\label{eq:Taylor expand}\\
    S_{2n}(\rho)
    &~=~\frac{1}{(2n)!}\frac{\partial^{2n}E(\rho,\delta)}{\partial\delta^{2n}}\bigg|_{\delta=0},\label{eq:various order symmetry energy}
  \end{align}
\end{subequations}
where $E_0(\rho)=E(\rho,\delta=0)$ denotes the EoS of symmetric nuclear matter, and the coefficients $S_{2n}(\rho)$ give the $2n$-order symmetry energy, where $n=1$ presents the density-dependent symmetry energy $S_2(\rho)$ and $n=2$ the fourth-order symmetry energy $S_4(\rho)$, respectively,
\begin{subequations}\label{eq:symmetry energy and fourth-order symmetry energy}
  \begin{align}
    S_2(\rho)&~=~\frac{1}{2!}\frac{\partial^2E(\rho,\delta)}{\partial\delta^2}\bigg|_{\delta=0},\label{eq:symmetry energy}\\
    S_4(\rho)&~=~\frac{1}{4!}\frac{\partial^4E(\rho,\delta)}{\partial\delta^4}\bigg|_{\delta=0}.\label{eq:fourth-order symmetry energy}
  \end{align}
\end{subequations}
The odd-order terms of the expansion are discarded in Eq. \eqref{eq:Taylor expand} owing to assuming the charge-independence of nuclear force and neglecting the Coulomb interaction in the infinite nuclear matter. The density slope parameter $L$ is used to reflect the density dependence of $S_2(\rho)$ at saturation density $\rho_0$, which is defined as
\begin{align}\label{eq:slope parameter}
  L=3\rho_0\frac{\partial S_2(\rho)}{\partial\rho}\bigg|_{\rho=\rho_0}.
\end{align}
In addition, the various order of terms in isospin asymmetry $S_{2n}(\rho)$ can be decomposed according to the separation of the EDF, namely,
\begin{subequations}\label{eq:decomposed S2n}
  \begin{align}
    S_{2n}
    &=~S_{2n,pot}~+~S_{2n,kin},\label{eq:S2n,pot+S2n,k}\\
    S_{2n,pot}
    &~=~\frac{1}{(2n)!}\frac{\partial^{2n}}{\partial\delta^{2n}}\frac{1}{\rho}\left.\left[\sum_\phi\left(\varepsilon_\phi^D~+~\varepsilon_\phi^E\right)\right]\right|_{\delta=0}\nonumber\\
    &~\equiv~S_{2n,pot}^D~+~S_{2n,pot}^E,\label{eq:S2n,pot}\\
    S_{2n,kin}
    &~=~\frac{1}{(2n)!}\frac{\partial^{2n}(\varepsilon_{kin}/\rho)}{\partial\delta^{2n}}\Bigg|_{\delta=0}\nonumber\\
    &~\equiv~S_{2n,kin}^M~+~S_{2n,kin}^p~+~S_{2n,kin}^D~+~S_{2n,kin}^E,\label{eq:S2n,k}
  \end{align}
\end{subequations}
where $S_{2n,pot}$ and $S_{2n,kin}$ denote the potential and kinetic part of $S_{2n}$. Additionally, $S_{2n,pot}^D$ ($S_{2n,pot}^E$) corresponds to the contributions from the Hartree (Fock) terms to $S_{2n,pot}$, while $S_{2n,kin}^M$, $S_{2n,kin}^p$, $S_{2n,kin}^D$ and $S_{2n,kin}^E$ define the corresponding contributions to $S_{2n,kin}$ from Eq. \eqref{eq:decomposed kinetic energy}, respectively.

\section{Results and Discussion}

In this work, the properties of nuclear fourth-order symmetry energy, and correspondingly its effects on several related quantities in neutron stars are studied in the CDF theory. Much attention is paid to the role of the Fock terms. The calculations are performed by using the RHF functionals PKA1\cite{Long2007PRC76.034314}, PKO1\cite{Long2006PLB640.150}, PKO2 and PKO3\cite{Long2008EPL82.12001}, in comparison with the RMF functionals PKDD\cite{Long2004PRC69.034319}, TW99\cite{Typel1999NPA656.331}, DD-ME1\cite{Niksic2002PRC66.024306} and DD-ME2\cite{Lalazissis2005PRC71.024312}. These functionals have been adopted in the description of nuclear matter and finite nuclei successfully, taking the advantage of the density-dependent meson-nucleon couplings, by which the medium effects of nuclear force in a nuclear many-body system are taken into account effectively. Notice that due to the limitation of the approach itself, the $\pi$ and $\rho$-tensor couplings are missing in four selected RMF functionals, while the RHF ones PKO1, PKO2, PKO3 contains the $\pi$ couplings, and PKA1 contains both. The applications of the RHF theory to the physics of nuclear matter and neutron stars have addressed essential role of the Fock terms, and one can find the details in Refs. \cite{Bouyssy1985PRL55.1731, Long2006PLB640.150, Sun2008PRC78.065805, Hu2010PLB687.271, Hu2010EPJA43.323, Miyatsu2012PLB709.242, Long2012PRC85.025806, Katayama2012AJ203.22, Miyatsu2015AJ813.135, Jiang2015PRC91.025802, Zhao2015JPG42.095101}. For the symmetry energies in Eq. \eqref{eq:various order symmetry energy}, the seven-point finite difference method is adopted in the practical calculations. The convergence and stability of the numerical results have been checked carefully, and the step size of isospin asymmetry $\delta$ is suggested as 0.01.

\subsection{Properties of nuclear fourth-order symmetry energy}

\subsubsection{Density dependence of nuclear fourth-order symmetry energy}
\begin{figure}[h]
  \centering
  \includegraphics[width=0.48\textwidth]{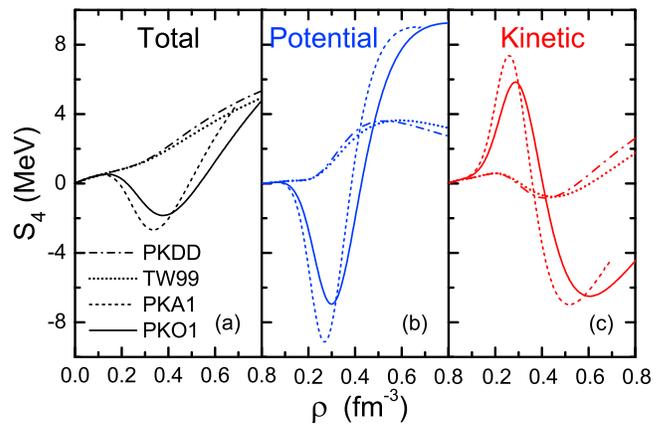}
  \caption{(Color online) The nuclear fourth-order symmetry energy $S_4$ and its potential part $S_{4,pot}$ and kinetic part $S_{4,kin}$ as functions of the baryonic density $\rho$. The results are calculated with RHF functionals PKA1 and PKO1, in comparison with RMF ones PKDD and TW99.}\label{Fig:S4}
\end{figure}

It is generally agreed that the effects of nuclear fourth-order symmetry energy $S_4(\rho)$ become non-negligible at high densities and at extreme isospin like in the interior of neutron stars, while its density dependent behavior is still poorly known. From the calculations of the selected CDF functionals, in Fig. \ref{Fig:S4} are shown the density dependence of $S_4(\rho)$. It is seen that all CDF models predict the similar curves at around and lower than the saturation density $\rho_0$, but the distinct deviation between the RHF and the RMF functionals occurs at the supranuclear density region. While the fourth-order symmetry energies in the RMF models increase monotonously with respect to the density $\rho$, those in the RHF turn to decrease beyond $\rho\gtrsim\rho_0$, even being negative when $\rho\thicksim[0.2, 0.5]$~fm$^{-3}$, and arise monotonously again after reaching the minimum value. The results are illustrated further by dividing the $S_4(\rho)$ into its potential part $S_{4,pot}$ and kinetic part $S_{4,kin}$, as shown in the middle and right panels of Fig. \ref{Fig:S4}, respectively. It is found that the density dependence of $S_{4,pot}$ and $S_{4,kin}$ becomes stronger in the RHF models than RMF ones. Specifically, within the RHF a clear reduction of $S_{4,pot}$ at $\rho\thicksim[0.1, 0.5]$~fm$^{-3}$ appears, which actually dominates the trend of the total fourth-order symmetry energy, leading to the values of $S_4(\rho)$ in RHF lower systematically than those in RMF.

\begin{figure}[t]
  \centering
  \includegraphics[width=0.48\textwidth]{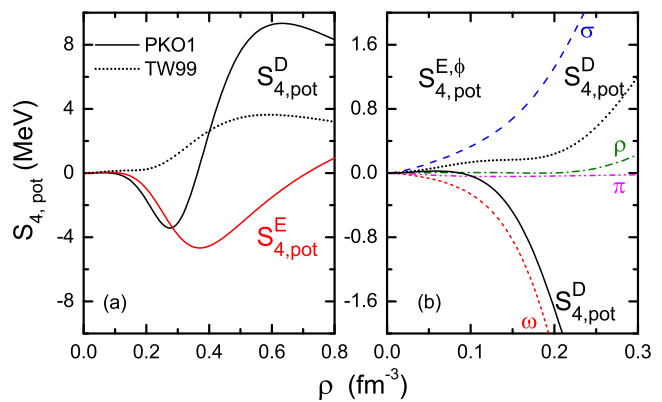}
  \caption{(Color online) (a) The potential part of nuclear fourth-order symmetry energy $S_{4,pot}$ is decomposed into the Hartree part $S_{4,pot}^D$ and the Fock part $S_{4,pot}^E$ according to Eq. \eqref{eq:S2n,pot}, as functions of the baryonic density $\rho$. The results are calculated with the RHF functional PKO1 (solid lines), in comparison with the RMF one TW99 (dotted line). (b) The Hartree part $S_{4,pot}^D$ and the contributions to $S_{4,pot}^E$ from the $\sigma$-, $\omega$-, $\rho$- and $\pi$-meson coupling channels are shown in detail at low densities.}\label{Fig:S4_p}
\end{figure}

To clarify the origin of the difference of $S_{4,pot}$ between two kinds of CDF models, it is convenient to take the results in Fig. \ref{Fig:S4}(b) apart into their contributions from the Hartree and Fock channels according to Eq. \eqref{eq:S2n,pot}, namely $S_{4,pot}^D$ and $S_{4,pot}^E$, as shown in Fig. \ref{Fig:S4_p}(a). For convenience, only the results from the RHF functional PKO1 and the RMF one TW99 are displayed, while the other functionals actually do not change the analysis and the conclusion. It is seen that the divergence on the density dependence of $S_{4,pot}$ results mainly from the Hartree part $S_{4,pot}^D$, which can be explained by the differences in the magnitude of meson-nucleon coupling constants and their density dependence between the RHF and RMF functionals. In fact, due to the extra interaction brought about by the Fock terms, the balance between the nuclear attractions and the repulsions is changed, which sequentially causes the difference in the coupling constants. Aside from the contribution of Hartree terms, the contribution from the Fock terms $S_{4,pot}^E$ can not be ignored as well, by giving a strongly suppressed contribution (maximum as $\backsimeq-4.7$~MeV) to $S_{4,pot}$ in a fairly broader density region. By separating $S_{4,pot}^E$ further in terms of meson-nucleon coupling channels, as plotted in Fig. \ref{Fig:S4_p}(b), one can find such an extra contribution from the Fock diagrams is mainly due to the isoscalar coupling channels, with a remarkably negative contribution from the $\omega$-meson coupling channel compensated by a positive contribution from the $\sigma$-meson one. Because the isovector meson-nucleon couplings decrease rapidly with respect to the density, as the case in present RHF functionals, the contributions to $S_{4,pot}^E$ from the $\rho$- and $\pi$-mesons are relatively weak. The results demonstrate again the importance of the isoscalar mesons via Fock diagram to the symmetry energy and correspondingly the essential role from the isospin-triplet components of the exchange potential EDF \cite{Zhao2015JPG42.095101}.

\begin{figure}[h]
  \centering
  \includegraphics[width=0.48\textwidth]{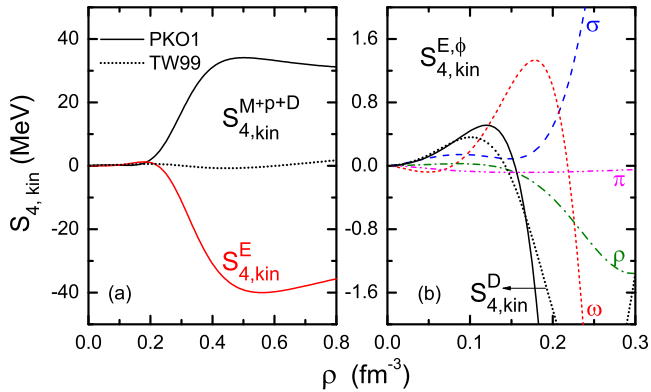}
  \caption{(Color online) (a) The kinetic part of nuclear fourth-order symmetry energy $S_{4,kin}$ is divided into various components according to Eq. \eqref{eq:S2n,k}, namely $S_{4,kin}^{M+p+D}=S_{4,kin}^M+S_{4,kin}^p+ S_{4,kin}^D$ and the Fock part $S_{4,kin}^E$, as functions of the baryonic density $\rho$. The results are calculated with the RHF functional PKO1 (solid lines), in comparison with the RMF one TW99 (dotted line). (b) The Hartree part $S_{4,kin}^D$ and the contributions to $S_{4,kin}^E$ from the $\sigma$-, $\omega$-, $\rho$- and $\pi$-meson coupling channels are shown in detail at low densities.}\label{Fig:S4_k}
\end{figure}

Similar as $S_{4,pot}$, the kinetic part of nuclear fourth-order symmetry energy $S_{4,kin}$ can be divided further according to Eq. \eqref{eq:S2n,k}, as shown in Fig. \ref{Fig:S4_k}(a). In the RHF calculation, $S_{4,kin}^{M+p+D}$ exhibits stronger density dependence than that in RMF when $\rho\gtrsim0.2~\rm{fm}^{-3}$, with a rapid growing up and then a very slow dropping down as the density increases. Such deviation between two functionals results from two aspects, for $S_{4,kin}^M+S_{4,kin}^p$ the $E^*$ plays a dominate role, while for $S_{4,kin}^D$ the role of the Dirac mass $M^*$ is partly involved, as seen in Eq. \eqref{eq:decomposed kinetic energy}. In fact, the starred quantity $E^*$ corresponds to the Landau mass in the CDF theory if the momentum dependence of the self-energy is left out. Therefore, it is the isospin dependence of the effective mass, including both the Dirac mass and the Landau mass, that accounts for the kinetic part of $S_4(\rho)$. Furthermore, extra contribution $S_{4,kin}^E$ from Fock diagrams is found, which provides a relatively small and positive value around the saturation density but an explicit suppression when $\rho\gtrsim0.2~\rm{fm}^{-3}$. From Fig. \ref{Fig:S4_k}(b), it is revealed that the density dependence of $S_{4,kin}^E$ is dominantly ascribed to the isoscalar $\omega$-meson coupling channel. Hence, the significant role of the Fock terms in the isospin properties of nuclear matter is demonstrated in both potential and kinetic part of the fourth-order symmetry energy.

\subsubsection{Nuclear fourth-order symmetry energy at saturation density}
To get compared with the constraints from other approaches, it is useful to discuss the properties of nuclear fourth-order symmetry energy $S_4(\rho)$ at saturation density $\rho_0$. Table \ref{Tab:S2 S4} shows the values of symmetry energies and their potential and kinetic components at $\rho_0$ with the different CDF functionals. It is worth noting that the values of $S_4(\rho_0)$ with the RHF functionals are systematically smaller than those in the RMF ones. While $S_4(\rho_0)$ predicted in RMF is located around $0.65~\rm{MeV}$, it is estimated to be about $0.35\thicksim0.58~\rm{MeV}$ in RHF, in which the functional PKA1, due to the inclusion of extra $\rho$-tensor coupling, gives the smallest value of $S_4(\rho_0)=0.352~\rm{MeV}$. Thus, it is expected that the involvement of the Fock terms in the CDF theory reduces the fourth-order symmetry energy at $\rho_0$, and the results are still in agreement with those from the density functional theory \cite{Chen2009PRC80.014322, Pu2017arXiv1708.02132, Cai2012PRC85.024302, Gonzalez2017arXiv1706.02736} and the chiral pion-nucleon dynamics \cite{Kaiser2015PRC91.065201, Wellenhofer2016PRC93.055802}, i.e., in general less than 2 MeV, but smaller in magnitude than the latest prediction by an extended nuclear mass formula \cite{Wang2017PLB773.62}.

\ifthenelse{\boolean{2column}}{
\begin{table*}
}{
\begin{sidewaystable}
}
  \centering
  \caption{Bulk properties of symmetric nuclear matter at saturation density $\rho_0$ (in unit of fm$^{-3}$), i.e., the symmetry energy $S_2(\rho_0)$, the density slope of symmetry energy $L$ (in unit of MeV), and the fourth-order symmetry energy $S_4(\rho_0)$ (in unit of MeV). $S_{2,pot}$ (or $S_{4,pot}$) and $S_{2,kin}$ (or $S_{4,kin}$) correspond to the potential part and the kinetic part of $S_2(\rho_0)$ (or $S_4(\rho_0)$), respectively. The results are calculated by using the RHF functionals PKA1, PKO1, PKO2 and PKO3, as compared to those given by the RMF functionals PKDD, TW99, DD-ME1 and DD-ME2.}
  \label{Tab:S2 S4}
  \setlength{\tabcolsep}{12.5pt}
  \begin{tabular}{ccccccccrc}
    \hline
    \hline
    Model&Interaction&$\rho_0$&$S_2(\rho_0)$&$S_{2,pot}$&$S_{2,kin}$&$S_4(\rho_0)$&$S_{4,pot}$&$S_{4,kin}$&$L$\\
    \hline
            &PKO1   &0.152   &34.370   &30.661   &3.709   &0.522   &-0.726   &1.248 &97.7 \\
    RHF     &PKO2   &0.151   &32.492   &28.094   &4.398   &0.583   &-0.510   &1.093 &75.9 \\
            &PKO3   &0.153   &32.987   &29.717   &3.270   &0.473   &-0.872   &1.345 &83.0 \\
            &PKA1   &0.160   &36.015   &35.551   &0.464   &0.352   &-1.770   &2.122 &103.5 \\
    \hline
            &DD-ME1 &0.152   &33.065   &24.692   &8.373   &0.649   & 0.170   &0.479 &55.5  \\
    RMF     &DD-ME2 &0.152   &32.295   &24.036   &8.259   &0.651   & 0.169   &0.482 &51.2  \\
            &TW99   &0.153   &32.767   &24.774   &7.993   &0.661   & 0.167   &0.494 &55.3  \\
            &PKDD   &0.150   &36.790   &28.657   &8.133   &0.645   & 0.168   &0.477 &90.2  \\
    \hline
    \hline
  \end{tabular}
\ifthenelse{\boolean{2column}}{
\end{table*}
}{
\end{sidewaystable}
}

By dividing into the potential and kinetic part, as shown in Tab. \ref{Tab:S2 S4}, the reduction of $S_4(\rho_0)$ in RHF can be explained by the fact that, although the kinetic parts $S_{4,kin}$ are enhanced, the potential parts $S_{4,pot}$ are sufficiently lowered than RMF. As has been discussed and illustrated in Fig. \ref{Fig:S4_p}(b), for $S_{4,pot}(\rho_0)$ the systematic deviation between two kinds of CDF models is attributed to both the difference of the Hartree part $S_{4,pot}^D$ and extra suppression from the Fock terms $S_{4,pot}^E$, specifically from the $\omega$-meson coupling channel. Besides, for the kinetic part $S_{4,kin}(\rho_0)$, the RHF models predict values of about $1.09\thicksim2.12~\rm{MeV}$, systematically larger than the selected density dependent RMF results. After extracting further the components of $S_{4,kin}(\rho_0)$ according to Eq. \eqref{eq:S2n,k}, as listed in Table \ref{Tab:S4_k}, it is clarified that in RHF the increase of $S_{4,kin}(\rho_0)$ at the saturation density results considerably from the Fock terms $S_{4,kin}^E$, especially in the $\omega$-meson coupling channel (see $S_{4,kin}^{E,\omega}$ in Fig. \ref{Fig:S4_k}(b)), since the contributions from the summation of the rest parts, namely
\begin{equation}
  S_{4,kin}^{M+p+D}=S_{4,kin}^M+S_{4,kin}^p+S_{4,kin}^D,
\end{equation}
are similar among all selected CDF models.

\begin{table}
  \centering
  \caption{Decomposition of the kinetic part of nuclear fourth-order symmetry energy $S_{4,kin}$ at saturation density $\rho_0$ according to Eq. \eqref{eq:S2n,k}, namely, the rest mass part $S_{4,kin}^M$, the momentum part $S_{4,kin}^p$, the Hartree part $S_{4,kin}^D$ and their summation $S_{4,kin}^{M+p+D}=S_{4,kin}^M+S_{4,kin}^p+S_{4,kin}^D$, as well as the Fock part $S_{4,k}^E$. The results are calculated with RHF functionals PKA1, PKO1, PKO2 and PKO3, and with RMF ones PKDD, TW99, DD-ME1 and DD-ME2. All values are in unit of MeV.}
  \label{Tab:S4_k}
  \setlength{\tabcolsep}{6pt}
  \begin{tabular}{crcrcc}
    \hline
    \hline
    Interaction&$S_{4,kin}^M$&$S_{4,kin}^p$&$S_{4,kin}^D$&$S_{4,kin}^{M+p+D}$&$S_{4,kin}^E$\\
    \hline
    PKO1    &-0.234  &0.408  &0.084   &0.258    &0.990 \\
    PKO2    &-0.647  &0.455  &0.332   &0.140    &0.953 \\
    PKO3    &-0.036  &0.374  &-0.010  &0.328    &1.017 \\
    PKA1    &1.667   &0.116  &-1.174  &0.609    &1.513 \\
    \hline
    DD-ME1  &-0.754  &1.212  &0.021   &0.479    &----- \\
    DD-ME2  &-0.698  &1.215  &-0.035  &0.482    &----- \\
    TW99    &-0.507  &1.233  &-0.232  &0.494    &----- \\
    PKDD    &-0.685  &1.203  &-0.041  &0.477    &----- \\
    \hline
    \hline
  \end{tabular}
\end{table}

Recently, from the $^{12}$C(e,e$^\prime$pN) scattering experiments at JLab, it is suggested that the protons and neutrons in a nucleus can form strongly correlated nucleon pairs, with large relative momentum, which are referred to as SRC pairs and regarded as a consequence of the nucleon-nucleon tensor force \cite{Sargsian2005PRC71.044615, Schiavilla2007PRL98.132501, Subedi2008Science320.1476}. Later on, it is realized that these SRC pairs could make significant influence on the kinetic part of nuclear symmetry energy \cite{Xu2012CPL29.122102, Hen2015PRC91.025803, Cai2016PRC93.014619}, and its fourth-order term $S_{4,kin}$ as well \cite{Cai2015PRC92.011601}. By using the Fermi gas model with correlated high-momentum neutron-proton pairs, a larger value $7.18\pm2.52~\rm{MeV}$ of $S_{4,kin}(\rho_0)$ with $\rho_0=0.16~\rm{fm}^{-3}$ is obtained in comparison with $\backsimeq0.45~\rm{MeV}$ predicted by the free Fermi gas model \cite{Cai2015PRC92.011601}. Besides, in the calculation within the CDF theory, it is also revealed that the inclusion of the Fock terms reduces sizably the kinetic part of $S_2$ at and above the saturation density\cite{Zhao2015JPG42.095101}, which is regarded partly as the effect of the nuclear tensor force \cite{Jiang2015PRC91.025802}. Here in this work, we demonstrate further the effects of the Fock terms on the kinetic part of $S_4$, especially from the isoscalar meson coupling channels, which correspondingly leads to the enhancement of $S_{4,kin}(\rho_0)$ in RHF systematically.

\subsection{Effects of nuclear fourth-order symmetry energy on neutron star properties}
The study of the fourth-order symmetry energy allows access to cold dense nuclear matter, such as that found in a neutron star. In particular when describing its cooling and rotational properties, the theoretical predictions could be affected essentially by the inclusion of $S_4$ in the EoS \cite{Steiner2006PRC74.045808, Xu2009PRC79.035802, Xu2009AJ697.1549, Cai2012PRC85.024302, Seif2014PRC89.028801}. Due to the extra contributions from the Fock terms in deciding the density dependence of $S_4$, the difference between RMF and RHF in the calculations of neutron star properties will be emphasized in the following discussions, namely in the proton fraction, the core-crust transition as well as the moment of inertia utilized to describe pulsar glitches.

\subsubsection{Proton fraction}
The cooling rate of neutron stars could be enhanced efficiently through the DUrca process, i.e. $n\rightarrow p+e^-+\bar{\nu}_e$ and $p+e^-\rightarrow n+\nu_e$, leading the star to cool off rapidly by emitting the thermal neutrinos \cite{Lattimer1991PRL66.2701}. The occurrence of the DUrca process rely sensitively on the proton fraction $\chi_p\equiv\rho_p/\rho$ of neutron star matter. For the description of neutrino free neutron star matter with the nucleons (neutrons and protons) and leptons (electrons and muons), the $\beta$-equilibrium, baryon density conservation and charge neutrality conditions are imposed here. The chemical potentials of nucleons and leptons satisfy the equilibrium conditions, constrained by the weak interacting reactions,
\begin{equation}
    \mu_\lambda=\mu_n-\mu_p,\label{eq:chemical equilibrium}
\end{equation}
where $\lambda = e^-, \mu^-$. The equations of motion for the leptons are the free Dirac equations. Therefore, the chemical potentials of leptons can be determined by the relativistic energy-momentum relation at the Fermi momentum,
\begin{align}
    \mu_{\lambda}=\sqrt{m_\lambda^2+(3\pi^2\rho\chi_\lambda)^{2/3}},\label{eq:mulepton}
\end{align}
where $m_\lambda$ denotes the lepton masses, $m_e=0.511$ MeV and $m_\mu=105.658$ MeV, respectively. The lepton fractions $\chi_\lambda\equiv\rho_\lambda/\rho$ in neutron star matter. When the chemical potential of electron $\mu_e$ reaches the threshold of the muon mass, the lepton $\mu^-$ will appear.

In order to extract the effects of symmetry energy on the proton fraction $\chi_p$, it is convenient to deduce the relation between nucleon chemical potentials by the thermodynamical relation, shown as
\begin{align}
\mu_n-\mu_p = 2\frac{\partial E(\rho,\delta)}{\partial\delta}.\label{eq:mun-mup}
\end{align}
Substituting the Eqs. \eqref{eq:mulepton}-\eqref{eq:mun-mup} into Eq. \eqref{eq:chemical equilibrium}, it is easily found that the lepton fraction $\chi_\lambda$ is actually the function of $E(\rho,\delta)$, and consequently the proton fraction $\chi_p$ is expressed as,
\begin{align}\label{eq:proton fraction}
  \chi_p(\rho) = \frac{1}{3\pi^2\rho}\sum_{\lambda}
  \left\{\left[2\frac{\partial E(\rho,\delta)}{\partial\delta}\right]^2 - m_\lambda^2\right\}^{3/2}.
\end{align}
deduced from the charge neutrality condition, $\chi_p = \chi_e + \chi_\mu$. Hence, by taking the Taylor series of expansion for $E(\rho,\delta)$ into account, given in Eq. \eqref{eq:Taylor expand}, and making an appropriate cut off to $n$, the influence of $2n$-order symmetry energy $S_{2n}$ on the proton fraction can be explored quantitatively.

\begin{figure}[h]
  \centering
  \includegraphics[width=0.48\textwidth]{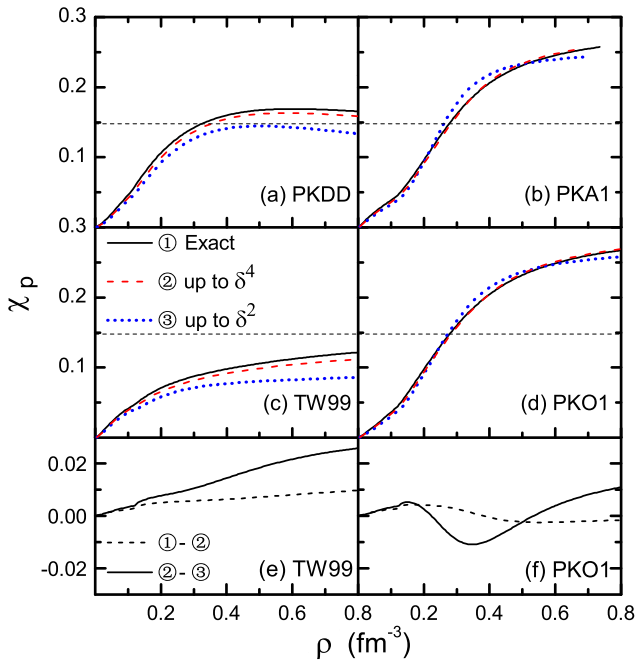}
  \caption{(Color online) The proton fraction $\chi_p$ as a function of the baryonic density $\rho$ in neutron star matter (panels (a)-(d)), calculated from Eq. \eqref{eq:proton fraction}, where $E(\rho,\delta)$ is taken as its exact values (\textcircled{1}, solid lines), approximated up to the fourth (\textcircled{2}, dashed lines) or the second (\textcircled{3}, dotted lines) order according to Eq. \eqref{eq:Taylor expand}, respectively. The results from the RHF functionals PKA1 and PKO1 are displayed, in comparison with the RMF ones PKDD and TW99. The horizontal dashed lines give the threshold~14.8\% for the occurrence of the DUrca process. For the functionals TW99 and PKO1, the effects of the fourth-order and the higher-order symmetry energies are illustrated in the panels (e) and (f), respectively, by showing the divergence of the results with different approximations to $E(\rho,\delta)$, labeled as \textcircled{2}$-$\textcircled{3} and \textcircled{1}$-$\textcircled{2}.}\label{Fig:proton fraction}
\end{figure}

Figure \ref{Fig:proton fraction} shows the density dependence of the proton fraction $\chi_p$ in neutron star matter. Compared with the results using the exact $E(\rho,\delta)$, approximation of $E(\rho,\delta)$ up to $\delta^2$ order (dotted lines) would generate appreciable errors which are relatively more distinct in RMF calculations (see panels (a), (c) and (e)) than in RHF (see panels (b), (d) and (f)). It is seen that in RMF the density dependence of $\chi_p$ is clearly underestimated with $E(\rho,\delta)$ up to $\delta^2$ order. The error is then partly diminished by introducing the $\delta^4$ order contribution into $E(\rho,\delta)$, namely taking the effect from the fourth-order symmetry energy $S_4(\rho)$ into account. However, with the approximation of $E(\rho,\delta)$ up to $\delta^2$ order, overestimation of the proton fraction around the density region $\rho\thicksim[0.2, 0.5]$~fm$^{-3}$ is observed in the RHF predictions, which is then almost compensated by including the $S_4(\rho)$ induced contribution in $E(\rho,\delta)$, as seen in Fig. \ref{Fig:proton fraction} (b) and (d) with dashed lines. In fact, such a systematical distinction between RMF and RHF calculations is correlated intimately with their divergence of the symmetry energy. Specifically, it is seen that the density dependence of the contribution of $S_4$ to the proton fraction, as plotted by the solid lines in Fig. \ref{Fig:proton fraction} (e) and (f), exhibits the same trend as those of $S_4(\rho)$ shown in Fig. \ref{Fig:S4}(a). In addition, the negative contribution of $S_4$ to $\chi_p$ in the density range of $\rho\thicksim[0.2, 0.5]$~fm$^{-3}$ within the RHF functional PKO1 ascribes mainly to the Fock terms, which consequently change the role of the fourth-order symmetry energy in deciding the matter distribution in neutron stars.

\subsubsection{Core-crust transition density}
The stability of matter in neutron stars are found to be sensitive to the density dependence of the symmetry energy as well, which decides the phase transition between nuclei and uniform matter and defines the core-crust interface of neutron stars \cite{Kubis2007PRC76.025801, Lattimer2007Phyreport442.109}. The baryonic number density of two coexisting phases corresponds to the so-called core-crust transition density $\rho_t$ that separates the liquid core from the inner crust in neutron stars. To estimate $\rho_t$, several dynamical methods, such as the random phase approximation (RPA), are used as a realistic treatment to determine the stability of the uniform ground state against cluster formation \cite{Ducoin2007NPA789.403, Xu2009AJ697.1549, Fattoyev2010PRC82.025810, Ducoin2011PRC83.045810, Piekarewicz2014PRC90.015803}. A simplification of the dynamical method, namely the thermodynamical method, is obtained at long-wavelength limit when the Coulomb interaction is neglected\cite{Xu2009AJ697.1549}. With this approximation, the core-crust transition properties of neutron stars are studied using a variety of nuclear effective
models and microscopic approaches \cite{Kubis2007PRC76.025801, Ducoin2011PRC83.045810, Seif2014PRC89.028801, Atta2014PRC90.035802} . It has been compared that the dynamical method predicts a slightly smaller transition density, about $0.005\thicksim0.015~\rm{fm}^{-3}$ lower, than the thermodynamical calculation \cite{Ducoin2011PRC83.045810}. Here the thermodynamical method is adopted for simplicity, since we mainly focus on the role of the fourth-order symmetry energy in the core-crust transition properties and the relevant contributions from the Fock terms. Thus, the stability of uniform $npe$ matter is required to obey the following inequalities, namely the intrinsic stability condition of any single phase,
\begin{subequations}\label{eq:intrinsic stability}
  \begin{align}
    -~\left(\frac{\partial P}{\partial v}\right)_{\mu_{np}}~>~0,\label{eq:intrinsic stabilityA}\\
    -~\left(\frac{\partial\mu_{np}}{\partial q_c}\right)_v~>~0,\label{eq:intrinsic stabilityB}
  \end{align}
\end{subequations}
where $P$ is the total pressure of neutron star matter, $v=1/\rho$ denotes the average volume per baryon, $\mu_{np}=\mu_n-\mu_p$ represents the difference between neutron and proton chemical potentials, and $q_c$ corresponds to the average charge per baryon. Here the finite size effects due to surface and Coulomb energies of nuclei are ignored. In addition, by introducing a density dependent thermodynamical potential $V_{ther}(\rho)$, the stability condition of Eq. \eqref{eq:intrinsic stabilityA} can be equivalently expressed as \cite{Seif2014PRC89.028801},
\begin{align}\label{eq:equivalent intrinsic stability}
  V_{ther}(\rho)
  ~\equiv&~2\rho\frac{\partial E(\rho,\delta)}{\partial\rho}
  ~+~\rho^2\frac{\partial^2E(\rho,\delta)}{\partial\rho^2}\nonumber\\
  &-~\left[\rho\frac{\partial^2E(\rho,\delta)}{\partial\rho\partial\delta}\right]^2\left/\frac{\partial^2E(\rho,\delta)}{\partial\delta^2}\right.~>~0,
\end{align}
which will be violated when the baryonic density decreases and reaches a threshold with $V_{ther}(\rho_t)=0$, subsequently determining the critical density $\rho_t$ for the core-crust transition. In the following discussion, similar to the analysis for the proton fraction, the influence of various order of symmetry energy on the core-crust transition, by adopting the corresponding approximation for $E(\rho,\delta)$, can be investigated quantitatively as well.

\begin{figure}[h]
  \centering
  \includegraphics[width=0.45\textwidth]{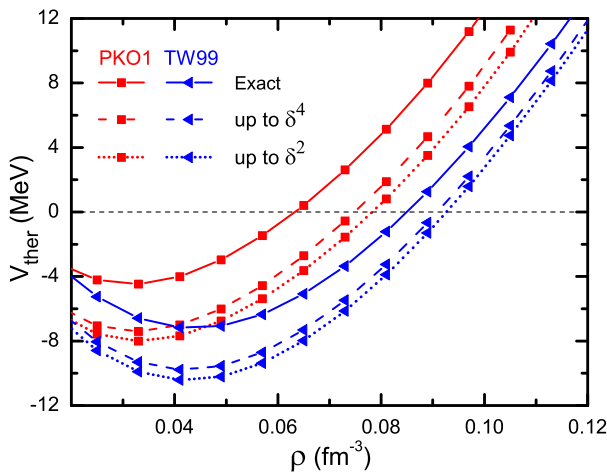}
  \caption{(Color online) The thermodynamical potential $V_{ther}$ as a function of the baryonic density $\rho$ in neutron star matter, calculated from Eq. \eqref{eq:equivalent intrinsic stability}, where $E(\rho,\delta)$ is taken as its exact values (solid lines), approximated up to the fourth (dashed lines) or the second order (dotted lines) according to Eq. \eqref{eq:Taylor expand}, respectively. The results are from the RHF functional PKO1, compared with the RMF one TW99.
  }\label{Fig:Vthermal}
\end{figure}

The density dependence of the thermodynamical potential $V_{ther}(\rho)$ is shown in Fig. \ref{Fig:Vthermal}, within the RHF functional PKO1 and the RMF one TW99 for comparison. Explicitly, the higher $S_{2n}$ terms within Eq. \eqref{eq:Taylor expand} the binding energy per nucleon $E(\rho,\delta)$ includes, the more enhanced the thermodynamical potential $V_{ther}(\rho)$ will be. By analyzing further the corresponding contributions from the three terms of $V_{ther}(\rho)$ as expressed in the rhs. of Eq. \eqref{eq:equivalent intrinsic stability}, it is then found that the isospin dependence of $E(\rho,\delta)$, namely the third term in Eq. \eqref{eq:equivalent intrinsic stability}, dominates such an enhancement of $V_{ther}(\rho)$ rather than its density dependence. Hence, the inclusion of these contribution from the high-order symmetry energy, such as from $S_4(\rho)$, results in the reduction of the core-crust transition density $\rho_t$. Moreover, in comparison with the curves from TW99, a systematical enhancement of the density dependence of $V_{ther}(\rho)$ is obtained within the RHF functional PKO1, no matter how the approximation on $E(\rho,\delta)$ is taken. As has been revealed, because of the extra contributions from the Fock terms, strong density dependence of the symmetry energy are predicted by the RHF calculations \cite{Sun2008PRC78.065805, Zhao2015JPG42.095101}, which naturally plays the role in the density dependence of $V_{ther}(\rho)$ via the first and the second terms in Eq. \eqref{eq:equivalent intrinsic stability}. As a result, it is expected that the inclusion of the Fock diagrams in the CDF theory reduces the core-crust transition densities $\rho_t$.

\ifthenelse{\boolean{2column}}{
\begin{table*}
}{
\begin{sidewaystable}
}
  \centering
  \caption{The density slope of symmetry energy $L$ (in unit of MeV), the core-crust transition density $\rho_t$ (in unit of $\rm{fm}^{-3}$), the corresponding values of the proton fraction $\chi_p$ and pressure $P_t$ (in unit of $\rm{MeV}\cdot\rm{fm}^{-3}$) at $\rho_t$ in neutron stars with the various CDF functionals. The results are obtained by adopting the different approximations to the thermodynamical potential Eq. \eqref{eq:equivalent intrinsic stability}, namely $E(\rho,\delta)$ is taken as its exact values (Exact), approximated up to the fourth (Quartic) or the second (Quadratic) order according to Eq. \eqref{eq:Taylor expand}, respectively.}
  \label{Tab:core-crust transition}
  \setlength{\tabcolsep}{5.2pt}
  \begin{tabular}{ccccccccccccccc}
    \hline
    \hline
    \multicolumn{1}{c}{\multirow{2}{*}{Interaction}}&\multicolumn{1}{c}{\multirow{2}{*}{$L$}}
    &\multicolumn{4}{c}{$\rho_t$}&\multicolumn{4}{c}{$\chi_p$}&\multicolumn{4}{c}{$P_t$}\\
    \cline{4-6}
    \cline{8-10}
    \cline{12-14}
    &&&EXACT&Quartic&Quadratic&&Exact&Quartic&Quadratic&&Exact&Quartic&Quadratic\\
    \hline
    PKO1     &97.7  &&0.0634  &0.0750  &0.0784  &&0.0219  &0.0263  &0.0276  &&0.3023  &0.4983  &0.5688 \\
    PKO2     &75.9  &&0.0745  &0.0805  &0.0824  &&0.0296  &0.0316  &0.0322  &&0.3449  &0.4371  &0.4694 \\
    PKO3     &83.0  &&0.0722  &0.0796  &0.0816  &&0.0278  &0.0302  &0.0308  &&0.3520  &0.4616  &0.4958 \\
    PKA1     &103.5 &&0.0550  &0.0670  &0.0701  &&0.0235  &0.0277  &0.0288  &&0.2567  &0.4083  &0.4553 \\
    \hline
    DD-ME1  &55.5 &&0.0843  &0.0917  &0.0939  &&0.0383  &0.0407  &0.0414  &&0.6040  &0.7230  &0.7616 \\
    DD-ME2  &51.2 &&0.0869  &0.0932  &0.0951  &&0.0388  &0.0406  &0.0411  &&0.5931  &0.6863  &0.7158 \\
    TW99    &55.3 &&0.0851  &0.0909  &0.0927  &&0.0368  &0.0387  &0.0393  &&0.5243  &0.6215  &0.6543 \\
    PKDD    &90.2 &&0.0755  &0.0866  &0.0902  &&0.0332  &0.0391  &0.0410  &&0.6142  &0.8836  &0.9869 \\
    \hline
    \hline
  \end{tabular}
\ifthenelse{\boolean{2column}}{
\end{table*}
}{
\end{sidewaystable}
}

In Tab. \ref{Tab:core-crust transition}, the core-crust transition densities $\rho_t$ of neutron stars with the different CDF functionals are given. The results are obtained by adopting the different approximations to the thermodynamical potential, namely $E(\rho,\delta)$ is taken as its exact values, approximated up to the fourth or to the second order according to Eq. \eqref{eq:Taylor expand}, respectively. In agreement with the analysis of $V_{ther}(\rho)$, it is revealed that the core-crust transition densities $\rho_t$ decrease monotonously for all models with the improvement of precision to describe the EoS, so do the corresponding proton fraction $\chi_p$ and the pressures $P_t$ at the density $\rho_t$. The same trend is found also in the recent studies from the other density functional approaches, either non-relativistic or relativistic \cite{Cai2012PRC85.024302, Seif2014PRC89.028801, Gonzalez2017arXiv1706.02736}. Therefore, it is claimed that the effects of the high-order symmetry energies are indispensable and the exact treatment of $E(\rho,\delta)$ is necessary in order to describe the core-crust transition properties appropriately. Furthermore, it should be noticed that the values of $\rho_t$, $\chi_p$ and $P_t$ given by RHF functionals are systematically smaller than those from the selected density dependent RMF, for instance, the exact values of $\rho_t\thicksim[0.055, 0.075]~\rm{fm}^{-3}$ within RHF while those of $\rho_t\thicksim[0.076, 0.087]~\rm{fm}^{-3}$ within the selected density dependent RMF, illustrating the essential roles of the Fock terms in the neutron star properties not only in the inner region but around the interface between core and crust.

\begin{figure}[h]
  \centering
  \includegraphics[width=0.46\textwidth]{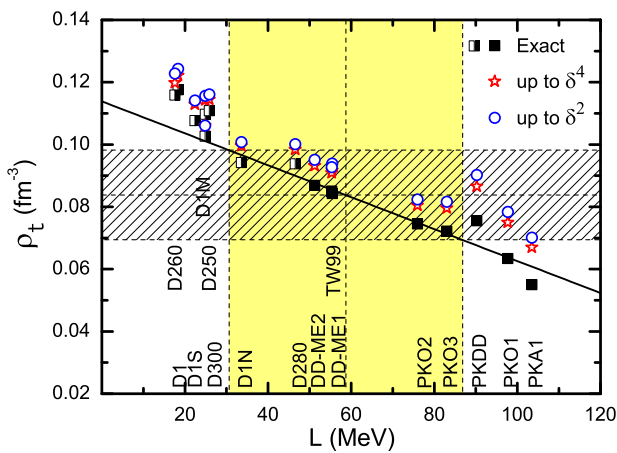}
  \caption{(Color online) Correlations between the core-crust transition densities $\rho_t$ of neutron stars and the density slope parameters $L$ of symmetry energies within the selected CDF functionals. The results within the Gogny density functionals taken from Ref. \cite{Gonzalez2017arXiv1706.02736} are plotted as well. From Tab. \ref{Tab:core-crust transition}, the transition densities $\rho_t$ are taken as its exact values (squares), approximations up to the fourth (empty stars) or the second order (empty circles), respectively. With the exact ones from the eight selected CDF models (solid squares), a linear fitting is given by the black solid line. The yellow (gray) region depicts the constraint on the density slope parameter $L=58.7\pm28.1~\rm{MeV}$ from Ref. \cite{Oertel2017RMP89.015007}, which determines further the constraint on the core-crust transition density with the thermodynamical method, namely $\rho_t\thicksim[0.069, 0.098]~\rm{fm}^{-3}$, based on the linear anti-correlation between $\rho_t$ and $L$, as marked by the shadowed area.}\label{Fig:rhot}
\end{figure}

As has been discussed, the density dependence of $V_{ther}(\rho)$ is enhanced in the RHF calculations because of the extra contributions to the symmetry energies from the Fock diagrams, leading correspondingly to the reduction of the core-crust transition densities $\rho_t$. While the density dependence of the symmetry energy is mainly reflected by its density slope parameter $L$, it is then worth investigating the relationship between $\rho_t$ of neutron stars and the density slope $L$ in nuclear matter, as illustrated in Fig. \ref{Fig:rhot}.

Similar to the previous discussion, three sets of $\rho_t$ are depicted respectively by adopting the different approximations to the thermodynamical potential $V_{ther}(\rho)$. For comparison, those from the Gogny density functionals are given as well \cite{Gonzalez2017arXiv1706.02736}. Systematically speaking, the RHF models predict larger values of $L$ compared to the Gogny and selected density dependent RMF ones, due to the effects of Fock terms \cite{Zhao2015JPG42.095101}. Noted that with RMF models which provide large value of $L$, such as several nonlinear RMF functionals \cite{Fattoyev2010PRC82.025810}, it is possible to provide smaller $\rho_t$. Thus, the small values for $\rho_t$ do not seem exclusively a consequence of the Fock diagrams, but are regards as a result that the models just have large density slopes $L$. Nevertheless, a linear anti-correlation is found approximately for the set of exact $\rho_t$ (squares), namely, the core-crust transition density decreases with the increasing density slope $L$, which is also satisfied for the cases that $\rho_t$ are approximated (stars or circles). Notice that such a $\rho_t$-$L$ correlation is found in several other studies as well \cite{Horowitz2001PRL86.5647, Xu2009AJ697.1549, Fattoyev2010PRC82.025810, Ducoin2011PRC83.045810, Moustakidis2012PRC86.015801, Providencia2014EPJA50.44, Gonzalez2017arXiv1706.02736}.

Utilizing the least-square method, it is then convenient to linearly fit these three sets of $\rho_t$-$L$ correlations, namely $\rho_t=aL+b$, where $a$ is in unit of $10^{-4}~\rm{fm}^{-3}\cdot\rm{MeV}^{-1}$, $b$ in $\rm{fm}^{-3}$ and $L$ in $\rm{MeV}$, respectively. To exhibit the linear anti-correlation in the CDF results, the fitting procedure is carried out only with the selected RMF and RHF functionals, dropping those with Gogny. Finally, there are $a = -5.13, -4.02, -3.73$, $b = 0.11$ and their Pearson correlation coefficients $r = -0.94, -0.89, -0.86$, respectively. Recently, the review study of the EoS for supernovae and compact stars collectively analyzes the impact of the nuclear symmetry energy and gives the constraint on the density slope parameter as $L=58.7\pm28.1$~MeV \cite{Oertel2017RMP89.015007}. Thus, proving the linear anti-correlation between $\rho_t$ and $L$ in collaboration with the empirical information on $L$, the constraint on the core-crust transition density is evaluated as $\rho_t\thicksim[0.069, 0.098]~\rm{fm}^{-3}$ (the shaded region in Fig. \ref{Fig:rhot}) if the exact value of $E(\rho,\delta)$ is used in determining the transition density $\rho_t$ with the thermodynamical method. There still exists a relatively large uncertainty of $\rho_t$, which may generate appreciable influence on the structure of neutron stars, specifically crucial for understanding the glitch phenomenon when they are rotating.

\subsubsection{Fraction of crustal moment of inertia}
As has been observed in many pulsars, the glitch phenomenon, i.e., the abrupt spin-up in the rotational frequency, is well believed to be the result of sudden transfers of angular momentum between the neutron superfluid permeating the inner crust and the rest of the star \cite{Anderson1975Nature256.25, Lattimer2007Phyreport442.109, Chamel2013PRL110.011101, Delsate2016PRD94.023008, Li2016ApJSS}. It is found that the rate of angular momentum transfer can be related to the fraction of the moment of inertia of the star which resides in the crust, as denoted by $\Delta I/I$ \cite{Link1999PRL83.3362}. Based on the slowly rotating assumptions for pulsars in the framework of general relativity \cite{Hartel1967APJ150.1005}, the fraction of crustal moment of inertia is well approximated by \cite{Lattimer2007Phyreport442.109, Lattimer2013AJ771.51}
\begin{align}\label{eq:moment of inertia}
  \frac{\Delta{I}}{I}~\simeq~\frac{8{\pi}P_tR^4}{3GM^2}~\left(\frac{MR^2}{I}~-~2\beta\right)~e^{-4.8\Delta{R}/R},
\end{align}
where $\beta=GM/(Rc^2)$ is the neutron star compactness parameter and $\Delta{R}/R$ denotes the crust thickness ratio. To compute the neutron star mass $M$, radius $R$ and crust thickness $\Delta{R}$ on the rhs. of Eq. \eqref{eq:moment of inertia}, the EoSs of neutron star matter under $\beta$-equilibrium as discussed above are used at high densities (neutron star core region), while BPS \cite{BPS1971ApJ...170..299Baym} and BBP \cite{BBP1971225BAYM} models are adopted to provide the EoS at low densities (neutron star crust region). The EoS of neutron star core and crust is matched by the core-crust transition pressure discussed before. The neutron star structure is then obtained by solving the stellar hydrostatic equilibrium equations, namely the Tolman-Oppenheimer-Volkov equations. Besides, the total moment of inertia of the star $I$ on the rhs. of Eq. \eqref{eq:moment of inertia} is estimated by \cite{Lattimer2005AJ629.979}
\begin{equation}
I\simeq(0.237\pm0.008)MR^2(1+2.84\beta+18.9\beta^4).
\end{equation}
As seen in Eq. \eqref{eq:moment of inertia}, it is mentioned that $\Delta{I}/I$ depends primarily on the stellar mass, radius and the pressure $P_t$ at the core-crust transition boundary, and scales as $P_tR^4M^{-2}$ \cite{Link1999PRL83.3362, Lattimer2013AJ771.51}.

\begin{figure}[h]
  \centering
  \includegraphics[width=0.45\textwidth]{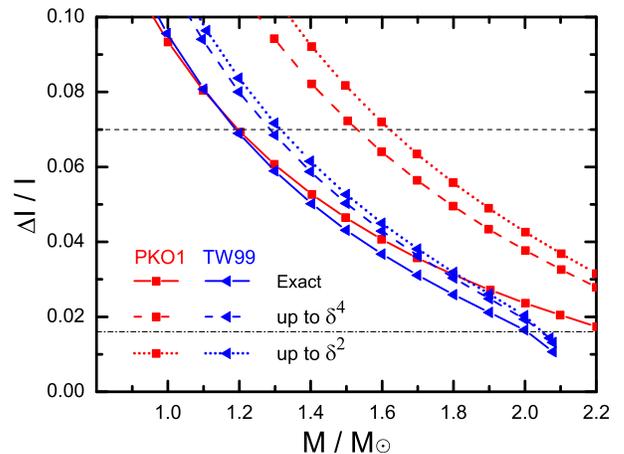}
  \caption{(Color online) The fraction of crustal moment of inertia $\Delta I/I$ as a function of the neutron star mass (in unit of the solar mass $M_\odot$). The results are calculated with the RHF functional PKO1 (lines with squares) and the RMF one TW99 (lines with triangles), by taking different sets of the core-crust transition pressure $P_t$ from Tab. \ref{Tab:core-crust transition}. Two horizontal lines represent the constraints on $\Delta I/I$, namely $\Delta I/I\geqslant0.016$ \cite{Andersson2012PRL109.241103} or $\Delta I/I\geqslant0.07$ \cite{Chamel2012PRC85.035801}, respectively.}\label{Fig:glitch}
\end{figure}

Figure \ref{Fig:glitch} shows the stellar mass dependence of the fraction of crustal moment of inertia of neutron stars based on different considerations of the core-crust transition pressure taken from Tab. \ref{Tab:core-crust transition}. It is revealed that, in both RHF and RMF results, $\Delta{I}/I$ decreases monotonously as the stellar mass goes up. When $E(\rho,\delta)$ reserves more the high-order components of $S_{2n}$, namely approaches gradually to its precise value, the pressure $P_t$ at the core-crust transition density will be brought down as listed in Tab. \ref{Tab:core-crust transition}. Subsequently, the suppression on the values of $\Delta{I}/I$ occurs mainly because of the reduction of transition pressure according to Eq. \eqref{eq:moment of inertia}, which becomes more remarkable for PKO1 since the pressure curves in RHF models stiffen further due to the contributions from the Fock terms. Although the values of $P_t$ within RHF are smaller systematically than those from the selected density dependent RMF, as shown in Tab. \ref{Tab:core-crust transition}, it is interesting to see that the curves of  $\Delta{I}/I$ with PKO1 and TW99 approach to each other in the case of exact calculations. In fact, the predicted radii of neutron stars in RHF are generally larger than those in density dependent RMF functionals in a wide range of stellar mass \cite{Sun2008PRC78.065805, Long2012PRC85.025806}. As a result, the counterbalance between the suppressed roles of the Fock terms in $P_t$ and the enlarged effects of the Fock terms on the radius $R$ takes place, making the exact calculations of $\Delta{I}/I$ less model dependent.

\ifthenelse{\boolean{2column}}{
\begin{table*}
}{
\begin{sidewaystable}
}
  \centering
  \caption{The maximum allowed neutron star masses within both the RMF and RHF functionals constrained by two sets of criterion from the measured glitches in Vela pulsar \cite{Andersson2012PRL109.241103}, based on the different consideration of the core-crust transition pressure taken from Tab. \ref{Tab:core-crust transition}.}
  \label{Tab:glitch}
  \setlength{\tabcolsep}{13.8pt}
  \begin{tabular}{cccccccccc}
    \hline
    \hline
    \multicolumn{1}{c}{\multirow{2}{*}{Interaction}}
    &\multicolumn{4}{c}{~$M/M_\odot$~~~~~~~~($\Delta{I}/I=0.016$)}
    &\multicolumn{4}{c}{~$M/M_\odot$~~~~~~~~($\Delta{I}/I=0.07$)}&\\
    \cline{3-5}
    \cline{7-9}
    &&Exact&Quartic&Quadratic&&Exact&Quartic&Quadratic \\
    \hline
    PKO1    &&2.25  &2.43  &2.45  &&1.20  &1.53  &1.62 \\
    PKO2    &&2.28  &2.39  &2.41  &&1.20  &1.36  &1.41 \\
    PKO3    &&2.35  &2.45  &2.47  &&1.24  &1.44  &1.49 \\
    PKA1    &&2.13  &2.35  &2.39  &&1.07  &1.36  &1.44 \\
    \hline
    DD-ME1  &&2.43  &2.45  &2.45  &&1.45  &1.59  &1.62 \\
    DD-ME2  &&2.46  &2.49  &2.49  &&1.45  &1.56  &1.59 \\
    TW99    &&2.01  &2.05  &2.06  &&1.19  &1.29  &1.31 \\
    PKDD    &&2.33  &2.33  &2.33  &&1.55  &1.78  &1.85 \\
    \hline
    \hline
  \end{tabular}
\ifthenelse{\boolean{2column}}{
\end{table*}
}{
\end{sidewaystable}
}

Finally, one could discuss more about the observational constraints on the fraction of crustal moment of inertia. The standard model for pulsar glitches holds that they are due to the neutron superfluid in the star's crust. In this case, the observed glitch rates and magnitudes for the Vela pulsar lead to the following constraint \cite{Andersson2012PRL109.241103}
\begin{equation}
\Delta{I}/I\gtrsim0.016.
\end{equation}
To satisfy this criterion, the maximum allowed neutron star masses are found to be larger than 2~$M_\odot$ for all selected functionals, as shown in Fig. \ref{Fig:glitch} and summarized in Tab. \ref{Tab:glitch}, which are in reasonable agreement with the measured large pulsar masses for J1614$-$2230 and J0348$+$0432 \cite{Demorest, Antoniadis1233232}. Recently, it is argued that due to entrainment of superfluid neutrons in the crust \cite{Andersson2012PRL109.241103, Chamel2013PRL110.011101}, one would have to enlarge the inferred lower limit to $\Delta I/I$ as
\begin{equation}
\Delta{I}/I\gtrsim0.07,
\end{equation}
in order to explain the measured glitches in Vela pulsar. For comparison, the maximum allowed neutron star masses in agreement with this constraint are listed in Tab. \ref{Tab:glitch}. It is seen that only very low mass neutron stars could satisfy this criterion, with the mass always lower than 2~$M_\odot$. It was thought that this new constraint would call the standard model for glitches into question \cite{Andersson2012PRL109.241103, Chamel2013PRL110.011101, Lattimer2013AJ771.51, Delsate2016PRD94.023008}, and the core-crust coupling during glitches would be necessary as one of the possible solutions \cite{Link2014AJ789.141, Newton2015MNRAS454.4400}. However, by taking pairing into account explicitly in the calculations of the effects of band structure on the neutron superfluid density in the crust of neutron stars, it is argued that the standard models of glitches based on neutron superfluidity in the crust can not be ruled out yet \cite{Watanabe2017PRL119.062701}. Nevertheless, the physics in pulsar glitches is still an open problem and need to be explored further, while in this work, attention is paid to the effects of nuclear high-order symmetry energies and the crucial influence of the Fock diagrams on the glitches related properties.

\section{Summary}
In this paper, by adopting the density dependent meson-nucleon coupling formalism, the density dependence of the nuclear fourth-order symmetry energy $S_4(\rho)$ and its properties at saturation density $\rho_0$ have been studied within the CDF theory. The calculations are performed by using the RHF functionals PKA1, PKO1, PKO2 and PKO3, in comparison with the RMF functionals PKDD, TW99, DD-ME1 and DD-ME2. It is found that the fourth-order symmetry energies $S_4(\rho)$ in RHF are considerably smaller than those in RMF at both saturation and supranuclear densities. It is illustrated then by analyzing the contributions from various meson-nucleon coupling channels to the potential and kinetic parts of $S_4(\rho)$. The studies clarify the important role of the Fock diagrams in determining the fourth-order symmetry energy, generally from three aspects. First, with the inclusion of the Fock terms, the density dependence and the magnitude of meson-nucleon coupling constants alternate, leading to divergent contribution of the Hartree terms of potential EDFs. Second, extra contributions are introduced by the Fock terms of potential EDFs, which is proved to be dominated by the isoscalar meson coupling channels. Third, the deviation in the potential EDFs affects the nucleon self-energies and changes consequently the kinetic EDFs via Eq. \eqref{eq:decomposed kinetic energy}. Quantitatively, the values of $S_4(\rho_0)$ at saturation density are estimated to be about $0.35\thicksim0.58~\rm{MeV}$ within selected RHF functionals, in consistence with several model predictions \cite{Chen2009PRC80.014322, Pu2017arXiv1708.02132, Cai2012PRC85.024302, Gonzalez2017arXiv1706.02736, Kaiser2015PRC91.065201, Wellenhofer2016PRC93.055802}, but smaller in magnitude than the latest one by an extended nuclear mass formula \cite{Wang2017PLB773.62}. Besides, the RHF models predict the values of the kinetic fourth-order symmetry energy $S_{4,kin}(\rho_0)$ about $1.09\thicksim2.12~\rm{MeV}$, systematically larger than the density dependent RMF results, which could be regarded partly as the effect of the nuclear tensor force embedded naturally in the exchange diagrams. One should notice here the comparison between RHF and RMF in this work is limited only to the density dependent meson-nucleon coupling version of the CDF approaches, so the conclusions may not be fulfilled by other CDF versions such as the nonlinear or point-coupling types of CDF approaches.

Furthermore, the effects of $S_4(\rho)$ on the neutron star properties have been investigated in detail, and the differences between RMF and RHF calculations are illustrated. To extract the contributions from $S_4$ and higher-order symmetry energies $S_{2n}$, the calculations are performed by taking the exact values of the equations of state or cutting them off up to the corresponding order in a Taylor series of expansion. Because of the suppressed roles in $S_4(\rho)$ brought about by the Fock diagrams, the fourth-order term $S_4(\rho)\delta^4$ of EoS within the RHF functionals actually contributes a negative value to determine the proton fraction $\chi_p$ in neutron star matter, which occurs in the density range of $\rho\thicksim[0.2, 0.5]~\rm{fm}^{-3}$. Besides, the fourth- and higher-order symmetry energies affect the core-crust transition properties as well, namely, reduce the core-crust transition density $\rho_t$ and the corresponding proton fraction as well as the transition pressure. With the inclusion of the Fock terms, the density dependence of the thermodynamical potential $V_{ther}$ becomes stronger as compared to the cases of the selected density dependent RMF. As a result, the core-crust transition densities $\rho_t$, the corresponding values of $\chi_p$ and $P_t$ are reduced further in RHF, demonstrating the essential roles of the Fock terms in the neutron star properties not only in the inner region but around the core-crust interface. In addition, a linear anti-correlation between the core-crust transition density $\rho_t$ and the density slope of symmetry energy $L$ is found in the CDF calculations in combination with the empirical information on $L$, which is then used to constrain the core-crust transition density as $\rho_t\thicksim[0.069, 0.098]~\rm{fm}^{-3}$. Noted that a simplified thermodynamical method has been used in this work to determine the properties of core-crust boundary. A more realistic dynamical method is then deserved along this line. Finally, it is also shown that the fraction of crustal moment of inertia of neutron stars is reduced by including the contributions of the high-order symmetry energies, in consistence with the behavior found in the transition pressure $P_t$.

In conclusion, the effects of the Fock terms in CDF approaches on nuclear isospin properties are demonstrated again in this work, which influence the nuclear fourth-order symmetry energy $S_4$ drastically, while $S_4$ is elucidated further to play the considerable role in understanding the physics of neutron stars, such as in cooling mechanism and pulsar glitches. Hence, the improvement in constraining the isospin and density dependence of the nuclear EoS with upcoming astrophysical observations and terrestrial experiments will be of utmost importance to interpret appropriately these topics with extreme physical conditions. The studies of these topics within a meson exchange picture of nuclear force, in particular with the inclusion of the Fock diagrams, are meaningful as well, for instance to answer what the role of the tensor force is. It is then expected that the RHF density functional will be improved effectively with the precise constraints on its isospin and density related properties.

\begin{acknowledgements}
This work is partly supported by the National Natural Science Foundation of China (Grant Nos. 11205075 and 11375076) and the Fundamental Research Funds for the Central Universities (Grant No. lzujbky-2016-30).
\end{acknowledgements}


\end{document}